\documentclass{kapedbk} % Computer Modern font calls

\usepackage{graphicx,amssymb}
\usepackage{dcolumn}% Align table columns on decimal point
\usepackage{bm}% bold math

\let\footnote\savefootnote

\normallatexbib

\begin{document}

%------------ article title  ------------------->>

%\articletitle[]{}

% If you use \\'s , please supply an alternate version of the title
% in square brackets, i.e.,
%\articletitle[Dialogo sui Massimi Sistemi]{Dialogo \\ sui Massimi Sistemi}

\articletitle[Isotropy of the velocity of light and the Sagnac
effect]{Isotropy of the velocity of light and the Sagnac effect}

%\articlesubtitle{and Last Recording Made by Hari Seldon}

%% optional, to supply a shorter version of the title for the running head:
\chaptitlerunninghead{Isotropy of the velocity of light}

%\author{}
%% multiple authors may be separated with \\
%% \author{Hari Seldon\\
%% and Golan Trevize}

%% or,

%% \author{Hari Seldon}
%% \author{Golan Trevize}
%% \and   % <=== Type in \and before the last author so that `and' will
%% \author{Eto Demerzel}    % print between the last two authors
                           % in the table of contents.

%\and %% <<== will make `and' appear in the Table of Contents. Use
     %%      before the last author is listed.

\author{J.-F. Pascual-S\'anchez\\A. San Miguel\\F. Vicente }
\affil{Dept. Matem\'atica Aplicada Fundamental\\ Secci\'on
Facultad
de Ciencias\\
    Universidad    de Valladolid\\ 47005, Valladolid, Spain}
    \email{jfpascua@maf.uva.es\\asmiguel@maf.uva.es\\fvicente@maf.uva.es}

\begin{abstract}
In this paper, it is shown, using a geometrical approach, the
isotropy of the velocity of light measured in a rotating frame in
the Minkowski space-time, and it is verified that this result is
compatible with the Sagnac effect. Furthermore, we find that this
problem can be reduced to the solution of geodesic triangles in a
Minkowskian cylinder. A relationship between the problems
established on the cylinder and on the Minkowskian plane is
obtained through a local isometry.
\end{abstract}


%------------ body of paper ------------------->>

\section{Introduction}\label{sec:1}
One of the most celebrated results of the Theory of Relativity is
the one known as the Sagnac effect \cite{sagnac}, which appears
when two photons describe, in opposite directions, a closed path
on a rotating disk returning to the starting point. Physically,
the Sagnac effect is essentially a phase shift between two
coherent beams of light travelling along paths in opposite senses
in an interferometer placed  on a rigidly rotating platform
\cite{post}. This phase shift can be explained as a consequence of
a time delay, so the Sagnac effect can also be measured with
atomic clocks timing light rays sent, e.g., around the rotating
Earth via the satellites of the Global Positioning System
\cite{ashby}. From a geometrical approach, such phase shift has
also been related \cite{anholonomity} to the fact that the time
component of the anholonomity object, corresponding to the choice
of an orthonormal frame on the space-time, is different from zero.

The Sagnac effect outlines the problem of the isotropy of the
velocity of light with respect to a non-inertial observer fixed on
the rotating disk. This problem has been treated from different
points of view. In \cite{selleri}, it is pointed out that the
Sagnac time delay, measured by one single clock, is due to an
anisotropy in the global speed of light for the non-inertial
observer, in contradiction with the local Einstein synchronization
convention. Another approach is found in \cite{rizzitar}. There,
the speed of light in opposite directions is the same, both
locally and globally. The proof is performed using three clocks
located at the initial and final positions of the two photons, and
by extrapolating point to point, the local Einstein
synchronization procedure to the whole periphery of the disk. The
disagreements between both approaches are connected with the
problem of the global time synchronization of points on the
periphery of a rotating disk. Only if this global synchronization
were possible there would exist a well defined spatial length
between different points on the boundary of the rotating disk.

In this paper we consider an ideal rotating disk with negligible
gravitational effects, thus the effects due to gravitational
fields ---that require the application of general relativistic
techniques as those in \cite{tar} or \cite{kopei}, where exact and
post-Minkowskian solutions are used--- are not considered here. We
will also show the isotropy of the velocity of light measured in a
rotating frame in the Minkowski space-time. We verify that this
isotropy is compatible with the Sagnac effect. For this we take
into account that every kinematical problem in special relativity
can always be translated into a geometrical problem on space-time.

Note on this respect that some authors have need to introduce some
dynamical explanations for explaining the rotating disk problem
\cite{phipps}.

An outline of the paper is as follows. In Sec. \ref{sec:2} we give
a brief account of the technique used by Rizzi and Tartaglia
\cite{rizzitar} and describe how the use of the {\sl hypothesis of
locality}, (see \cite{mash}) offers an explanation of the Sagnac
effect in the framework of special relativity, without using the
anisotropy of a global speed of light. In Sec. \ref{sec:3}, we
solve this problem in terms of the world-function associated to
the geodesic determined by the world-lines of the observer and the
photon and the simultaneity space corresponding to the observer.
In Sec. \ref{sec:4} a formulation of the problem using the
solution of geodesic triangles is obtained. Finally, in Sec.
\ref{sec:5} a relationship between the problem stated on a
Minkowskian cylinder and on a Minkowskian plane is obtained.

\section{The rotating disk and the Sagnac effect}\label{sec:2}
Let $\mathcal{D}\subset\mathbb{R}^3$ be a disk of radius $\rho$,
and let us denote by $\partial \mathcal{D}$ the circle bounding
$\mathcal{D}$. We consider an inertial reference frame $F:
(O^\prime,\{\bm{e}_i\}_{i=1}^3)$, where $O^\prime$ is the center
of $\mathcal{D}$ and  $\{\bm{e}_i\}_{i=1}^3$ is an orthonormal
basis for the Euclidean space $\mathbb{R}^3$. In the coordinate
system $(x,y,z)$ associated to $F$, the points in $\mathcal{D}$
have coordinate $z$ equal to zero. It will also be useful to
consider polar coordinates $(r,\theta)$ on $\mathcal{D}$. Now we
assume that the disk $\mathcal{D}$ is uniformly rotating about the
$O^\prime$ axis, with angular velocity $\omega$. In the space-time
$(\mathcal{M},\bm{\eta})$ of Special Relativity in Minkowski
coordinates, with $\bm{\eta}={\rm diag}\,(-1,-1,-1,c^2)$, the
motion of the points $P\in\partial \mathcal{D}$ with polar
coordinates $(\theta,t)$, is given by world-lines
$\gamma_{_P}:t\mapsto \gamma_{_\theta}(t)$, that in coordinates
$(x,y,z,t)$ can be expressed as
    \begin{equation}\label{1}
    \gamma_{_\theta}(t):\;(\rho\cos(\omega t+
    \theta),\,\;\rho\sin(\omega t+ \theta),\,\; 0,\,\; t).
    \end{equation}
This congruence of time-like curves determines a cylinder
$\mathcal{C}\subset\mathcal{M}$. On the cylinder $\mathcal{C}$
both a metric $\bm{g}$ is induced by the metric $\bm{\eta}$, which
in comoving coordinates $(\theta,t)$, reads
    \begin{equation}\label{4}
    \bm{g}=-\rho^2d\theta ^2 -2\, \omega \rho^2dt d\theta +
    \alpha^2(\omega)c^2\,dt^2,
    \end{equation}
where
    \begin{equation}\label{4b}
    \alpha^2 (\omega):=1-(c^{-1}\omega \rho)^2,
    \end{equation}

and a Killing vector field $\bm{\Gamma}$  given by a combination
of a rotation and a time translation, that, at each point
$P=(\theta,t)$, is $\bm{\Gamma}(P)=\dot{\gamma}_{_P}(t)$, are
defined. The associated  Killing congruence  has non null
vorticity within the cylinder but is zero outside it. So, the
vorticity and the 4-velocity play an analogous role to the
magnetic field and the 4-electromagnetic potential, respectively,
of the Aharonov-Bohm effect in electrodynamics, \cite{anandan}.
The metric (\ref{4}) is globally stationary and locally static;
therefore a {\it local} splitting of $\mathcal{C}$ can be obtained
using local hypersurfaces locally orthogonal to the trajectory of
the rotating observer, as in \cite{landau}. Even a {\it global}
operational quotient space by the Killing congruence can be build,
by using the radar distance as a spatial distance \cite{bel}.

In general, for every two points $A, B$, joined by a geodesic
$\gamma(u)$, being $u$ a special parameter with $\gamma(u_1)=A,
\gamma(u_2)=B$, there is a function $\Omega(AB)$
---the world function in Synge's terminology, \cite{Synge}---
defined by
    \begin{equation}\label{2803031}
    \Omega(A\,B):={\textstyle\frac{1}{2}}\!\int_{u_1}^{u_2}
    \bm{g}(\bm{v},\bm{v})\,ds
    \end{equation}
where $\bm{v}=\dot{\gamma}(u)$ denotes the tangent vector to the
geodesic $\gamma(u)$. Let us now consider, at the time $t=t_1$, a
point $O\in\partial \mathcal{D}$ and the world-line
$\gamma_{_O}(t)$ corresponding to a curve in the congruence
(\ref{1}), with $\theta=0$. On $\gamma_{_O}(t)$ one may build a
field of non-inertial reference frames $F^\prime(t)$. The proper
time interval between two events $P_0,P_1$ with coordinate times
$t_1$ and $t_2$ measured by the observer $F^\prime$ is given in
terms of the world-function (\ref{2803031}) as
    \begin{equation}\label{135}
    \tau_2-\tau_1:=c^{-1}\sqrt{2\Omega(P_0P_1)}=
    \alpha(\omega)(t_2-t_1).
    \end{equation}

Suppose that the rotating observer fixed on the circle
$\partial\mathcal{D}$ carries a device which emits, at the time
$t=0$, two photons in opposite directions along the periphery of
the disk. The world-lines of both photons are null helices. Their
 equations in the inertial reference frame $F$ read
    \begin{equation} \label{3001}
    \gamma_{_{L\pm}}(t): \; \left (x_{_L}= \rho\cos(\pm\varpi t),\;
    y_{_L} = \rho\sin(\pm\varpi t), \;t=t\right),
    \end{equation}
where $\pm\varpi$ denotes the angular speeds of the photons given
by $\varpi \rho=c$, being the plus (resp. minus) sign associated
to the photon moving in the same (resp. opposite) sense as the
rotating disk.

At the initial time $t=0 $ it is assumed that $\gamma_{_O}(0)=
\gamma_{_L}(0) $. The world-line corresponding to each photon cuts
the curve $\gamma_{_O}$ at times $t^*_\pm$, for which it is
satisfied the condition $\gamma_{_L}(t_\pm)=\gamma_{_O}(t_\pm)$.
Therefore one obtains
    \begin{equation}\label{time}
    t^*_\pm=\frac{2 \pi} {\varpi\mp\omega}.
    \end{equation}

The relationship between proper time on $\gamma_{_O}$ and the
inertial coordinate time given in (\ref{135}) establishes that the
proper time in $F^\prime$ runs slow with respect to an inertial
one. Hence, using (\ref{time}) one obtains
    \begin{equation}\label{8}
    \tau_\pm=2\pi\frac{\alpha(\omega)}{\varpi\mp\omega}.
    \end{equation}
The proper time increment measured by the observer $F^\prime$
among the arrival times of the two photons $P_1=\gamma_{_O}(t_1)$
and $P_2=\gamma_{_O}(t_2)$ is (see, e.g. \cite{anandan}):
    \begin{equation}\label{9}
    \tau_+ - \tau_-=\frac{4\omega S}{c^2}\alpha^{-1}(\omega)
    \end{equation}
that, in the limit of small rotational speeds, takes the classical
form given in \cite{sagnac}:
    \begin{equation}\label{10}
    \tau_+ - \tau_- = \frac{4\omega S}{c^2}+O(c^{-4})
    \end{equation}
where $S=\pi r_0^2$ is the disk area.

The Sagnac time delay is  the desynchronization  of a pair of
clocks after a complete round trip, which has been  initially
synchronized and  sent by the rotating observer to travel in
opposite directions, \cite{rizzitar}. In this case, the time
differences along a complete round trip on the periphery $\partial
\mathcal{D}$ of the disk, are not uniquely defined and the
measurement of each one must be corrected by half the Sagnac time
delay when compared with an identical clock remaining fixed at the
initial position.  After this correction is made, the global light
speed is the same for the photon moving on $\partial\mathcal{D}$
in opposite sense. This is in fact what is done in the Global
Positioning System, \cite{ashby}. Note, on the other hand, that if
the readings of both clocks are not corrected by half the Sagnac
time (\ref{9}), one obtains an anisotropic velocity of light, as
in \cite{selleri}.

\section{Measurement of relative speeds in Minkowski space-time}\label{sec:3}
Let us assume that at the time $t=0$, in the inertial reference
frame $F$, a non-inertial observer $F^\prime$ at a fixed point
$P_0\in\partial \mathcal{C}$ emits a pulse of light in the same
direction of the movement of the disk. The event $P_0$ corresponds
in the cylinder $\mathcal{C}$ to the point with cylindrical
coordinates $(\theta,t) =(0,0)$. We now determine the relative
speed of the ray of light with respect to the non-inertial frame
$F^\prime$.

The world-lines for the observer and the photon can be expressed
in cylindrical coordinates, in the form
    \begin{equation}
    \gamma_{_O}(t):\; (\theta = 0,\; t=t),\qquad \gamma_{_L}(t):\;
    (\theta = (\varpi - \omega)t,\;t=t )
    \end{equation}
respectively. The curve $\gamma_{_O}(t)$ is the time-like helix
corresponding to a non-inertial observer fixed at the point $O$ on
the disk. The curve $\gamma_{_L}(t)$ describes the null helix of
the photon co-rotating with the disk.

On the world-line $\gamma_{_O}$ one can determine a vector field
$\bm{\Lambda}$ such that the orthogonality condition,
$\bm{g}(\bm{\Lambda},\dot{\gamma_{_O}}(t))=0$, is satisfied. For
each point $P$ on $\gamma_{_O}$, one builds a space-like geodesic
$\gamma_{_S}$ on the cylinder $(\mathcal{C},\bm{g})$,
corresponding to the initial data $P,\bm{\Lambda}(P)$
    \begin{equation}
    \gamma_{_S}:\; \left(\theta =
    (\varpi^2-\omega^2)\,\omega^{-1}\,(t-t_0),\; t=t\right).
    \end{equation}
The geodesic $\gamma_{_S}$ can be interpreted as the locus of
locally simultaneous events on an arc of the circle $\partial C$
as seen by the rotating observer. For the construction of the
simultaneity space $\gamma_{_S}$, the hypothesis of locality given
in \cite{mash} is used, which establishes the local equivalence of
an accelerated frame and a local inertial frame with the same
local speed. In this way, a slicing of the cylinder $\mathcal{C}$
through a family of sections $\gamma_{_S}$ orthogonal to the
congruence of curves $\gamma_{_P}$ is obtained.

\begin{figure}[h]\centering{
    \includegraphics[scale=1,angle=0]{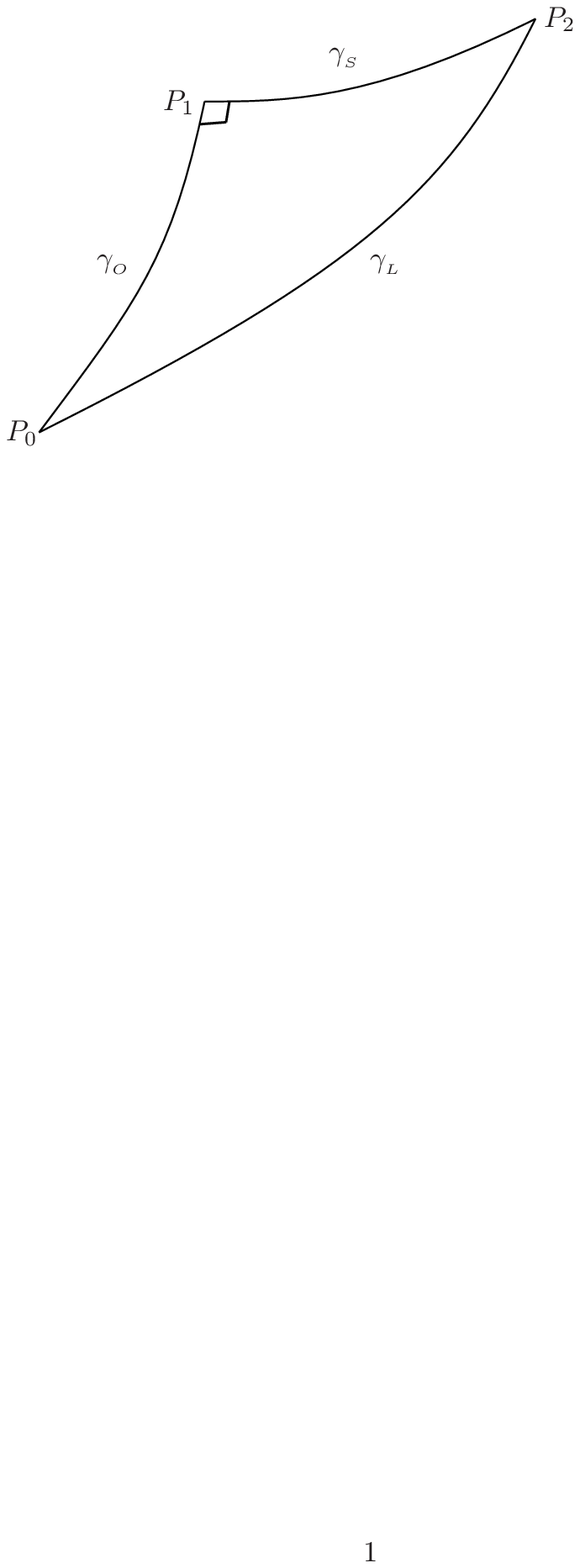}}% Here is how to import EPS art}
    \caption{\label{fig1}
        Geodesic triangle on the cylinder $(\mathcal{C},\bm{g})$.
        $\gamma_{_O}, \gamma_{_L}$ denote, respectively, the world-line of
        the observer and the photon. $\gamma_S$ represents the
        simultaneity space at the point $P_1$.}
    \end{figure}

When the rotating frame reaches the point $P_1=(0,t_1)$ in the
curve $\gamma_{_O}$, the world-function $\Omega(P_0P_1)$ is the
square of the proper time (up to a constant factor) between the
events $P_0$ and $P_1$ of space-time, measured by the non-inertial
observer. At the time $t_1$ the photon lies on the point $P_2$ of
the local simultaneity space relative to the non-inertial rotating
observer. Since both the non-inertial observer and the photon move
on $\partial \mathcal{C}$, the corresponding points in space-time
remain on the Minkowskian cylinder $\mathcal{C}$. Consider the
point $P_2$ given by the intersection of the curves $\gamma_{_S}$
and $\gamma_{_L}$, (see Fig.~\ref{fig1}):
    \begin{equation}\label{11}
    P_2=\left((\varpi^2-\omega^2)\,\varpi^{-1}\,t_0,\;
    (\varpi+\omega)\,\varpi^{-1}\,t_0\right).
    \end{equation}

Using the metric (\ref{4}), the world-function corresponding to
the pairs of points $(P_0,P_1)$ and $(P_1,P_2)$, calculated along
the curves $\gamma _{_O}$ and $\gamma _{_S}$, are
    \begin{equation}\label{12}
    \Omega(P_0P_1)=-\Omega(P_1P_2)=ct_0\,\alpha(\omega).
    \end{equation}
respectively. Therefore, taking into account (\ref{135}), the
relative speed of the photon with respect to the non-inertial
frame defined by
    \begin{equation}\label{14}
    v_{_{L,O}}^2:=-c^2 \, \, \frac{\Omega(P_1P_2)}{\Omega(P_0P _1)},
    \end{equation}
coincides with $c^2$.

\section{Equivalent formulation of the problem }\label{sec:4}
\begin{figure}[h]\centering{
    \includegraphics[scale=1 ,angle=0]{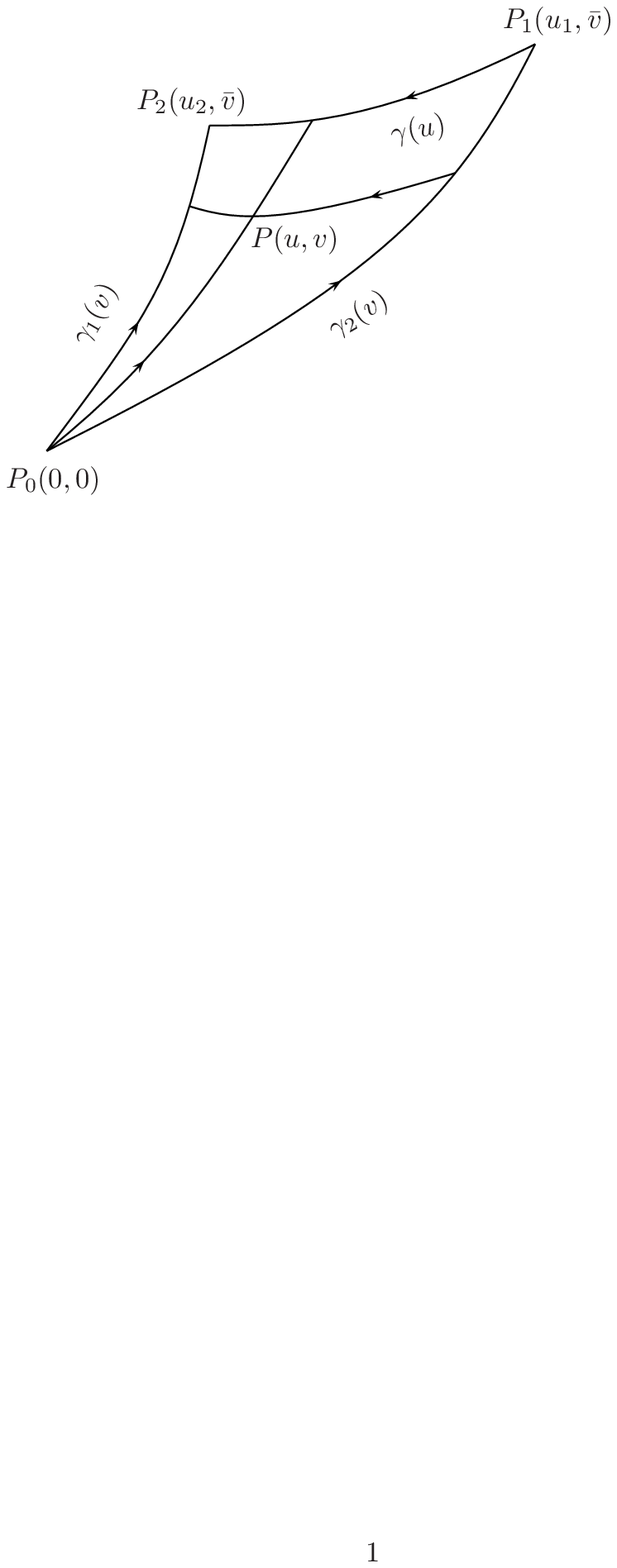}}% Here is how to import EPS art
    \caption{\label{fig2}
        Geodesic triangle on a surface with small curvature.
        The family of curves $\gamma(v)$ emanating from the
        point $P_0$, are geodesics parametrized by $u\in [0,\bar{v}]$.
        Transversal curves are geodesics parametrized by $u\in [u_1,u_2]$.}
    \end{figure}

Result (\ref{14}) can be compared with that obtained by using the
solution of geodesic triangles on the semi-Riemannian manifold
$(\mathcal{S},\bm{g})$ For this we consider a geodesic triangle
$P_0P_1P_2$ on a 2-manifold $(\mathcal{S},\bm{g})$ as shown in
Fig.~\ref{fig2}.  For arbitrary points $A$ and $B$, let
$\Omega_a(AB)$ denote the covariant derivative of (\ref{2803031})
with respect to the coordinates of $A$, and denote by
$\Omega^a(AB)$ the vector associated to $\Omega_a(AB)$ by means of
the metric $\bm{g}$.

Let us assume that the Riemannian curvature of a surface
$\mathcal{S}$ is small and we will use the same notation as in
\cite{Synge}, Chapter II. If $\{\lambda_0(P_0),\lambda_1(P_0)\}$
is an orthonormal basis on $T_{P_0}\mathcal{S}$, one can build a
field of reference frames on $\mathcal{C}$ by parallel transport
of this frame along all geodesics $x^i(v)$ through $P_0$. On the
field $\{\lambda_0(P),\lambda_1(P)\}$, the vector field
$V^i:=\partial x^i/\partial v$, tangent to one of these geodesics
on an arbitrary point, has constant components
$V^{(a)}=V^i\lambda_i^{(a)}$. On the other hand, the components of
the symmetrized Riemann tensor
    \begin{equation}\label{3103031}
    S_{ijkl}:=-{\textstyle\frac{1}{3}}(R_{ijkm}+R_{imjk}),
    \end{equation}
will be denoted by $S_{(abcd)}$.

For the geodesic triangle determined by the curves
$P_0P_1:\gamma_1(v)$, $P_0P_2:\gamma_2(v)$ and $P_1P_2:\gamma(u)$
(with $u\in[u_1,u_2], v\in [0,\bar{v}]$, see Fig.~\ref{fig2}) a
relationship between the world-functions of the sides of this
triangle is obtained in \cite{Synge}:
    \begin{equation}\label{4000}
    \Omega(P_1\,P_2)=\Omega(P_0\,P_1)+\Omega(P_0\,P_2)-
    \Omega_{a}(P_0\,P_1)\Omega^a(P_0\,P_2)+\phi,
    \end{equation}
where
    \begin{equation}\label{2803032}
    \phi:={\textstyle\frac{1}{6}}\!\int_0^{\bar{v}}(\bar{v}-v)^3D^4_v\Omega\,dv
    \end{equation}
and $D^4_v\Omega$ denotes the covariant derivative of fourth order
of the world-function $\Omega(\gamma_1(v),\gamma_2(v))$ for an
arbitrary $v\in[0,\bar{v}]$. An explicit approximate expression
for $\phi$ appears in \cite{Synge} p. 73, written in terms of the
Riemann tensor and its covariant derivatives. An application of
this solution to build Fermi coordinates in general space-times of
small curvature is given in \cite{GRSV}. In general it is
satisfied that
    \begin{equation}\label{3130032}
    \phi_0=\frac{3}{(u_2-u_1)^3}\int_0^{\bar{v}}\int_{u_1}^{u_2}
    q(u,v)\,[1122]\, du\, dv,
    \end{equation}
where $q(u,v)$ is the polynomial
    \begin{equation}
    q(u,v):=(\bar{v}-v)^3\big((u_2-u)^2+(u-u_1)^2\big),
    \end{equation}
and symbol $[1122]$, defined as
    \begin{equation}\label{3103033}
    [1122]:=-{\textstyle\frac{1}{3}}S_{(a_1b_1c_2d_2)}V^{(a_1)}
    V^{(b_1)}V^{(c_2)}V^{(d_2)},
    \end{equation}
is constant on $\mathcal{S}$, so that $\phi_0$ vanishes. In
(\ref{3103033}) $V^{(i_1)},V^{(i_2)}$ are the components of $V$ at
points $P_1,P_2$ respectively. In the problem considered in this
work, the metric (\ref{4}) is uniform on the cylinder
$\mathcal{C}$, and the Riemannian curvature is zero, therefore
expression (\ref{2803032}) vanish.

Therefore, one obtains for the solution of the same triangle in
the point $P_1$
    \begin{equation}\label{5000}
    \Omega(P_0\,P_2)=\Omega(P_1\,P_0)+\Omega(P_1\,P_2)-
    \Omega_{a}(P_1\,P_0)\Omega^a(P_1\,P_2),
    \end{equation}
where the  covariant derivatives  are calculated now at the point
$P_1$. Now, since the geodesic $P_0\,P_2$ is null and  the
geodesics $P_1\,P_0$ and $P_1\,P_2$  are orthogonal in $P_1$;
then, from (\ref{5000})  one obtains
    \begin{equation} \label{6000}
    \Omega(P_1\,P_2)=-\Omega(P_0\,P_1)
    \end{equation}
Consequently, the ratio
    \begin{equation}\label{2803034}
    v_{_{L,O}}^2:=-c^2\;\frac{\Omega(P_1\,P_2)}{\Omega(P_0\,P_1)},
    \end{equation}
coincides with (\ref{14}).

\section{Reduction to the Minkowskian plane}\label{sec:5}
In this section, we will  see that the rotating observer on the
disk has a specific characteristic which other different
non-inertial observers do not have in general. In the first place,
it is observed that expression (\ref{135}), which relates the
proper time $\tau $ of a non-inertial observer fixed on the
rotating disk (moving with constant angular speed $\omega$, such
that $\omega \rho=v $) to the coordinate time $t$, coincides with
the expression relating the inertial observer's time to the time
of another inertial reference frame boosted with rectilinear speed
$v$. Then one concludes that only by measuring proper time, a
rotating observer will not be able to determine the local inertial
or non-inertial character of the frame rotating uniformly on the
disk. The only magnitude that he will be able to measure in that
case is the speed modulus $v$.

\begin{figure}[h]\centering{
    \includegraphics[scale=1,angle=0]{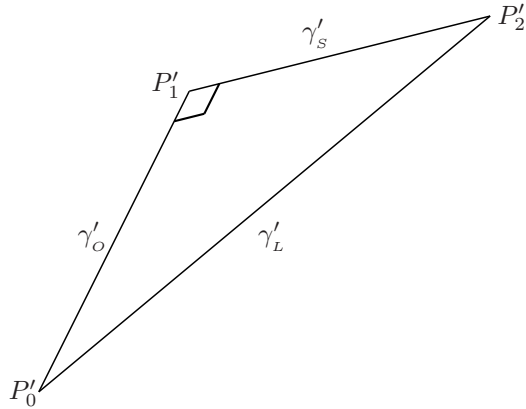}}% Here is how to import EPS art
    \caption{\label{fig3}
        Geodesic triangle on the Minkowskian plane
        $(\mathcal{P},\bm{\eta})$. $\gamma_{_O}^\prime,\gamma_{_L}^\prime$
        denote, respectively, the world-line of the observer and
        the photon. $\gamma_S^\prime$ represents the simultaneity
        space at the point $P_1^\prime$.}
    \end{figure}

Now, let us consider a boosted rectilinear inertial frame $K$. To
measure the speed of a photon moving in the same direction as $K$
with respect to this frame we consider the configuration shown in
Fig.~\ref{fig3}.

\noindent Here $\gamma_{_O}^\prime$ represents the straight line
described by the observer (we are assuming that the speed is $v $)
in a Minkowskian plane $(\mathcal{P}, \bm{\eta})$. On the other
hand, the null straight line $\gamma_{_L}^\prime$ represents the
trajectory that one photon describes, and, finally, the line
$\gamma_{_S}^\prime$ is the space-like straight line of
simultaneous events
 to the emission event of the photon. This line is
everywhere $\eta$--orthogonal to the observer line at the event
$P^{\prime}_1:(vt_0,t_0) $. Explicitly, taking $P^\prime_0=(0,0)$,
these curves are given by
    \begin{equation}
    \gamma_{_L}: (x=ct, t), \qquad \gamma_{_S}:
    \big(x=v^{-1}c^2(t-t_0\alpha^2(v)), t\big)
    \end{equation}
where now $\alpha^2(v):=1-v^2/c^2$. This can be verified directly
from Figure 2.

Moreover, point $P^{\prime}_2 $ at which $\gamma_{_S}^\prime$ cuts
to $\gamma_{_L}^\prime$ has the coordinates
    \begin{equation}\label{15}
    P^\prime_2: \left (\,t_0(c+v),\, c^{-1}t_0(c+v)\,\right).
    \end{equation}
Therefore, keeping in mind again that $P^{\prime}_1=(vt_0,t_0) $,
one obtains that the distances between $P^\prime_1$ and
$P^\prime_2$ along $\gamma^\prime_{_L}$ and between $P^\prime_0$
and $P^\prime_1$ along $\gamma^\prime_{_O}$ are
    \begin{equation}\label{17}
    -\tilde{\Omega}(P^{\prime}_1P^{\prime}_2)=
    \tilde{\Omega}(P'_0P'_1)=\alpha(v)ct_0
    \end{equation}
where $\tilde{\Omega}(AB)$ denotes the world-function associated
to points $A,B$ and the metric $\bm{\eta}$. The relative speed
between the light ray  and the boosted rectilinear inertial
observer, defined through the ratio
    \begin{equation}\label{19}
    v_{_{L,O^\prime}}^2=-c^2\,\frac{\tilde{\Omega}(P^{\prime}_1P^{\prime}_2)}
    {\tilde{\Omega}(P^{\prime}_0P^{\prime}_1)},
    \end{equation}
coincides with $c^2$.

The identity between these expressions and those obtained before
in  Sec. \ref{sec:3} is clear. Indeed, if $v$ is substituted for
$\omega \rho$ those expressions are coincident. The fact that the
values of $\tilde{\Omega}(P^\prime_0P^\prime_1)$ and
$\tilde{\Omega}(P^\prime_1P^\prime_2)$ coincide with the values
$\Omega(P_0P_1) $ and $\Omega(P_1P_2)$ obtained in the problem
solved on the cylinder is due to a local isometry between the
Minkowskian plane $(\mathcal{P}, \bm{\eta})$, which contains the
line of universe of the boosted rectilinear inertial observer, and
the Minkowskian cylinder $(\mathcal{C}, \bm{g})$, which contains
the world-line of the non-inertial rotating frame.

As pointed out at the beginning of this section, the non-inertial
rotating observer on the disk has a specific characteristic which
other different non-inertial observers do not have, in general. In
this case, the expression (\ref{135}), relating the proper time
$\tau $ and the coordinate time $t$, is the same as in inertial
frames. This allows to build an isometry between cylinder
$\mathcal{C}$ and the plane $\mathcal{P}$ as follows.

Let $\phi:{\mathcal U}\subset {\mathcal C} \rightarrow {\mathcal
P}$ be a smooth map between a neighborhood of $\mathcal{U}$, which
contains the  geodesic triangle considered above, and the plane
$\mathcal{P}$. Denote by $T_P\mathcal{C}$ and
$T_{\phi(P)}\mathcal{P}$ the tangent spaces to $\mathcal{C}$ and
$\mathcal{P}$ at the points $P,\phi(P)$ respectively.  The map
${\phi}$ is such that its differential, $d\phi\, : T_P\mathcal{C}
\rightarrow T_{\phi (P)} \mathcal{P}$, is a linear isometry for
every point $P\in \mathcal{U}$:
    \begin{equation}\label{3000}
    \bm{\eta}(d\phi \left(\bm{v}),d\phi(\bm{w})\right)=\bm{g}(\bm{v},\bm{w}),
    \end{equation}
for every $\bm{v},\bm{w}\in T_P\mathcal{C}$. Let us consider a map
$\phi$ such that $\phi(t,\theta)= (t,x(t,\theta))$. We determine a
function $x(t,\theta)$ satisfying condition (\ref{3000}). This
function is determined through the partial differential system
    \begin{equation}
    \frac{\partial x}{\partial t}=\omega \rho, \quad \frac{\partial
    x}{\partial \theta}=\rho,
    \end{equation}
whose solution is the function $x(t,\theta)=\rho(\omega
t+\theta)$. Therefore an isometry as
    \begin{equation}\label{3103034}
    \phi\, : (t,\theta) \longmapsto \left(t,\rho(\omega
    t+\theta)\right),
    \end{equation}
maps $(\mathcal{C};\bm{g})$ into $(\mathcal{P};\bm{\eta})$,
retaining the same coordinate time  in both manifolds.

The geodesic triangle of vertices $P_0,P_1,P_2$ in $\mathcal{C}$
is mapped into the straight triangle $P_0^\prime,P_1^\prime,
P_2^\prime$ in $\mathcal{P}$. Therefore, it is possible to
translate the problem of measuring the speed of light with respect
to a non-inertial reference frame, which describes a circumference
rotating uniformly, to the problem of measuring the speed of light
by an inertial reference frame, being the velocity equal to $c$ in
both cases. By means of this local isometry, for the point $P_2$
on the cylinder there exists a corresponding $P^\prime_2$ in the
plane, which has the same coordinates as the event $P_2 $,
obtained in Sec. \ref{sec:3} by means of the hypothesis of
locality with the slicing of $\mathcal{C}$.

Returning to the initial problem of two photons describing the
periphery of a rotating disk in opposite senses, it is observed
that one obtains the same result for both photons, as it may be
verified solving the corresponding problem on the Minkowskian
plane, where the speed of light is independent of the direction
followed by the photons.

\section{Concluding remarks}\label{sec:6}
In \cite{rizzitar}, using the locus of locally simultaneous events
to the non-inertial rotating observer (given by space-like helices
in a Minkowski space-time), it is shown that the speed of light
measured by a non-inertial observer fixed on the disk rim always
turns out to be $c$ both locally and globally. The local isometry
(\ref{3103034}), shows how this coincidence  is obtained. In fact,
this local isometry allows to calculate relative speeds
(\ref{2803034}) and (\ref{19}) in the problem of the rotating
disk, mapping the problem from the multiply connected Minkowskian
cylinder to another one established in the simply connected
Minkowskian plane.

From the above reasoning, one observes that although the observer
is non-inertial this is not reflected on the measurements of
relative speeds. This is because the module of the centripetal
acceleration of the observer $\omega^2 \rho\,\alpha^{-2}(\omega)$,
coincides with the module of the normal curvature of the
world-line of the observer on the cylinder $(\mathcal{C},
\bm{g})$. A rotating observer corresponds to a Killing trajectory,
so its world-line is a geodesic on this cylinder. Moreover, the
Gaussian curvature of the cylinder is zero. So, the
non-inertiality of the rotating observer is not reflected in the
measurement procedure, because this  is only based on the first
fundamental form of $\mathcal{C}$.

Finally, we remark that the frame of reference considered in the
problem of a rotating disk is very special, so the problem can be
established on a circular cylinder. A more general case would be,
for example, that of a deformable closed loop filament moving and
preserving a non-circular shape.

\begin{acknowledgments}
The authors wish to thank A. Tartaglia for a discussion on the
subject of this work. This work was completed with the support of
the Junta de Castilla y Le\'on (Spain), project VA014/02.
\end{acknowledgments}


\newpage

\begin{chapthebibliography}{99}

\bibitem{sagnac}
G. Sagnac, {\it C.R. Acad. Sci. Paris}, {\bf 157}, 708 (1913).
\bibitem{post}
    E.J. Post, {\it Rev. Mod. Phys.}, {\bf 39}, 475, (1967);
    G.E. Stedman, {\it Rep. Prog. Phys.}, {\bf 60}, 615 (1997).
\bibitem{ashby}
    D. Allan, N. Ashby and M. Weiss, {\it Science}, {\bf 228}, 69 (1985).
\bibitem{anholonomity}
    J.F. Corum, {\it J. Math. Phys.}, {\bf 18}, 770 (1977);
    J.-F. Pascual-S\'anchez, Summaries of workshop D.1 at 14th
    International Conference on General Relativity and
    Gravitation, Florence, 1995, (unpublished).
\bibitem{selleri}
    F. Selleri, {\it Found. Phys. Lett.}, {\bf 10}, 73 (1997);
    A. Peres, {\it Phys. Rev. D}, {\bf 18}, 2173 (1978).
\bibitem{rizzitar}
    G. Rizzi and A. Tartaglia,  {\it Found. Phys.}, {\bf 28}, 1663
    (1998).

\bibitem{tar}
    A. Ashtekar, {\it J. Math. Phys.}, {\bf 16}, 341 (1975);\,
    A. Tartaglia, {\it Phys. Rev. D}, {\bf 58}, 064009  (1998).
\bibitem{kopei}
    S. Kopeikin and G. Sch\"{a}fer, {\it Phys. Rev. D}, {\bf 60}, 124002
(1999).
\bibitem{phipps}
G. Cavalleri, {\it Nuovo Cimento}, {\bf 53 B}, 415 (1968); \O.
Gr\o n,  {\it  Am. J. Phys.}, {\bf 43}, 869 (1975), {\it Found.
Phys.}, {\bf 10}, 499 (1980);  T. E. Phipps, {\it Found. Phys.},
{\bf 10}, 289 (1980), {\bf 10}, 811 (1980); D. Dieks, {\it Eur. J.
Phys.},{\bf \ 12 }, 253 (1991); T.A. Weber, {\it Am. J. Phys.},
{\bf 65}, 946 (1997).
\bibitem{mash} B. Mashhoon, {\it Phys. Lett. A}, {\bf 145}, 147 (1990);\,
    A. Tartaglia, in {\it Reference Frames and Gravitomagnetism}, eds. J.-F.
    Pascual-S\'anchez, L. Flor\'{\i}a, A. San Miguel, F. Vicente,
    (World Scientific, Singapore, 2001).
\bibitem{anandan}  J. Anandan, {\it Phys. Rev. D } {\bf 24}, 338
(1981).
\bibitem{landau}
    L. Landau and E. Lifshitz, {\it The Classical Theory of
    Fields}, (Pergamon Press, Oxford, 1971).
\bibitem{bel} Ll. Bel, J. Mart{\'{\i}}n, A. Molina, {\it J. Phys. Soc.
 Japan}, {\bf 63}, 4350 (1994); G. Rizzi, M. L. Ruggiero, {\it Found. Phys.}, {\bf 32},
 1525 (2002).
 \bibitem{Synge}
    J.L. Synge, {\it Relativity: The General Theory}, (North--Holland, Amsterdan,
    1960).
\bibitem{GRSV}
    J.M. Gambi, P. Romero, A. San Miguel and F. Vicente, {\it Int. J.
    Theor. Phys.}, {\bf 30}, 1097 (1991).

\end{chapthebibliography}
\end{document}